# ASTA AT FERMILAB: ACCELERATOR PHYSICS AND ACCELERATOR EDUCATION PROGRAMS AT THE MODERN ACCELERATOR R&D USERS FACILITY FOR HEP AND ACCELERATOR APPLICATIONS*

V. Shiltsev[#], P. Piot, Fermilab, Batavia, IL 60510, USA

## Abstract

We present the current and planned beam physics research program and accelerator education program at Advanced Superconducting Test Accelerator (ASTA) at Fermilab.

## ASTA FACILITY

The Advanced Superconducting Test Accelerator (ASTA) at Fermilab will enable a broad range of beam-based experiments to study fundamental limitations to beam intensity and to develop transformative approaches to particle-beam generation, acceleration and manipulation. ASTA incorporates a superconducting radiofrequency (SRF) linac coupled to a photoinjector and small-circumference storage ring capable of storing electrons or protons. ASTA will establish a unique resource for R&D towards Energy Frontier facilities and a test-bed for SRF accelerators and high-brightness beam applications. The unique features of ASTA include: (1) a high repetition-rate, (2) one of the highest peak and average brightness within the U.S., (3) a GeV-scale beam energy, (4) an extremely stable beam, (5) the availability of SRF and high-quality beams together, and (6) a storage ring capable of supporting a broad range of ring-based advanced beam dynamics experiments. These unique features will foster a broad program in advanced accelerator R&D and accelerator education which cannot be carried out elsewhere. Several experiments and test can be carried out at ASTA simultaneously at three experimental areas – see Fig.1. Detail description of ASTA can be found elsewhere [1,2]

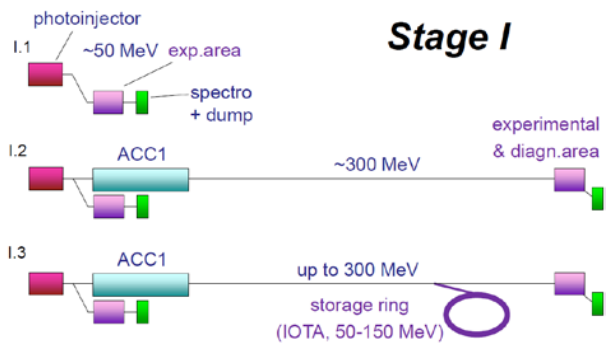

Fig.1: Schematics of ASTA (Stage I configuration).

## ASTA ACCELERATOR R&D THRUSTS

*Accelerator R&D for Particle Physics at the Intensity and Energy Frontiers*

The combination of a state-of-the-art superconducting linear accelerator and a flexible storage ring enables a broad research program directed at the particle physics accelerators of the future. The corresponding research program includes a number of proposals:

i) Test of non-linear, integrable, accelerator lattices (the Integrable Optics Test Accelerator) which have the potential to shift the paradigm of future circular accelerator design
ii) Test of space charge compensation in high intensity circular accelerators
iii) Advanced phase space manipulations
iv) Test of optical stochastic cooling
v) Flat-beam-driven dielectric-wakefield acceleration in slab structures
vi) Investigation of acceleration and cooling of carbon-based crystal structures for muon accelerators
vii) Measurement of the electron wave function size in a storage ring
viii) Studies of beam echo phenomena in a storage ring
ix) Development of a quadrupole pickup and study of the quadrupole mode dynamics in space-charge dominated beams
x) High power targetry studies for LBNE
xi) A tagged photon beam at ASTA for Detector R&D
xii) HEP applications of inverse Compton scattered photons.

It was specifically noted at the 2013 Community Summer Study ("Snowmass-2013" [2]) that there is obvious lack of accelerator test facilities where many burning accelerator science and technology issues of the Intensity Frontier (IF) accelerators [4] can be addressed. ASTA will provide this so-much needed platform for the IF accelerator R&D.

*Accelerator R&D for Future SC RF Accelerators*

High gradient, high power SC RF systems are critical for many accelerator facilities under planning for the needs of high-energy physics, basic energy sciences and other applications. ASTA offers a unique opportunity to explore the most critical issues related to SC RF technology and beam dynamics in SC RF cryomodules, such as:

i) Demonstration of High Power High Gradient SRF Cryomodules with Intense Beams
ii) Demonstration of technology and beam parameters for the Project X pulsed linac
iii) Wakefield measurements in the SC RF cryomodules
iv) Demonstration of ultra-stable operation of SC RF cryomoduyles with beam-based feedback
v) Development and experimental test of neural network based control systems and optimization tools for SRF accelerators
vi) Exploration of CW mode of operation of the ILC-type SRF cryomodule

*Accelerator R&D for Novel Radiation Sources*

High energy (up to 1 GeV), high peak and average power and high brightness electron beams have tremendous potential for generation of high-brightness high energy photon beams spanning the range from keVs to dozens of MeV. The high average power and brightness of the ASTA electron beam has unmatched potential for development of several novel radiation source ideas, such as:

i) High brightness X-ray channeling as a compact x-ray radiation source
ii) Inverse Compton scattering gamma-ray source and its applications
iii) Demonstration of feasibility of an XUV FEL oscillator
iv) Production of narrow-band gamma-rays
v) Laser-induced microbunching studies with high micropulse-repetition rate electron beams

*Accelerator R&D for Stewardship and Applications*

With its high energy, high brightness, high repetition rate, and the capability of emittance manipulations built-in to the facility design, ASTA is an ideal platform for exploring novel accelerator techniques of interest for very broad scientific community beyond high energy physics. The proposed research program includes several proposals and expressions of interest for such explorations:

i) Demonstration of techniques to generate and manipulate ultra-low emittance beams for future hard X-ray free-electron lasers
ii) Beam energy dechirper using corrugated pipes
iii) Development and test of a beam-beam kicker
iv) Coherent diffraction radiation measurements of bunch length

All the proposals can be found in [1] and have been discussed by the Fermilab's Accelerator Advisory Committee (February 6-8, 2013). Later, all of them have been presented at the 1st ASTA User's Meeting and reviewed by the ASTA Program Advisory Committee (July 23-24, 2013).

## ACCELERATOR EDUCATION OPPORTUNITIES AT ASTA, ASTA AND IARC

Accelerators have taken an increasingly important role in our Society and improving their performances, downsizing their footprints, or reducing their cost will be a long and challenging endeavour that will strongly rely on future generation of accelerator scientists. In addition, the multidisciplinary nature of Accelerator Science makes accelerators ideal platforms for student education over a vast range of topics.

New beam physics issues associated with future accelerators have emerged (single-bunch and multi-bunch collective effects in high brightness electron accelerators, simulations of large emittance beams in heavy ions and/or multi state charge accelerators) and more intricate accelerator physics tools capable of accounting for all these effects are being developed and need to be carefully validated at available facilities. Such models are mandatory to efficiently design and test the performance of these accelerators prior to their construction. Similarly, future accelerators will require the development of precise diagnostics and controls capable of measuring and/or correcting unprecedented beam parameters (sub-micrometer size, femtosecond duration, with high average current). Most of these topics will be explored to some extent at ASTA and could form the basis of research projects for PhD, Masters, or internships.

Interdisciplinary research projects involving the engineering discipline will also be possible at ASTA. Examples that come to mind are fabrication of state-of-the-art accelerator components requiring precision modeling, drafting and manufacturing. RF engineering R&D examples are ubiquitous in accelerator physics. Likewise, novel synchronization techniques and feedback systems will be needed, e.g. to insure that the accelerating fields in the superconducting linear accelerators are very stable.

Over the last decade the number of Educational Institutions that have taken on developing a curriculum in Accelerator Science has increased from a couple to about 10 [5]. None of these Universities currently operates a full-fledged accelerator complex and generally rely on facilities available at National Laboratories to carry most of their experiments. National laboratories have developed programs to foster partnership with Universities. Fermilab was actually the first laboratory to initiate such a selective program, which offers to financially support students to carry-out Accelerator-Science research at Fermilab's facilities [6]. This program – the Joint University - Fermilab Doctoral Program in Accelerator Physics and Technology – has graduated close to 40 students since its initiation in 1987. More than half of the program's graduates have remained in the field of Accelerator Science, and approximately one quarter of them are prominent scientists with leadership positions in national laboratories or universities.

Fermilab also created prestigious fellowships to attract promising scientists to carry-out research in Accelerator Science: the Peoples fellowshipwas created in 2001 and the Bardeen-Engineering Leadership fellowship [7] was created in 2005. The John Bardeen Engineering Leadership Program is designed to provide entry-level opportunities for outstanding engineering graduates who are interested in working in a cutting edge research environment. Fermilab provides opportunities in the fields of electrical, electronics, radio frequency systems, power distribution, magnets, RF cavities, mechanical, materials science and cryogenic engineering. The Peoples Fellowship was created at Fermilab with the goal of attracting outstanding accelerator scientists early in their careers, both to enhance Fermilab's capabilities in accelerator science and related technologies, and to train and develop the accelerator scientists and technologists who will carry our field forward in the future. Together these fellowship programs have attracted and trained approximately 20 people within the last decade. Most of these fellows still carry-out research in Accelerator Science.

Finally, Fermilab has also been active in involving younger scientists (undergraduate students) through various summer programs. One of these programs, the Lee Teng Undergraduate Internship in Accelerator Science and Engineering established by Fermilab and Argonne, aims at attracting undergraduate students into the world of particle accelerator physics and technology. The selected students first attend a two-week general accelerator physics class at the summer US particle accelerator school thereby gaining a broad overview of Accelerator Science. They then work on a research project for the following weeks under the supervision of Fermilab and Argonne staff members. ASTA offers the possibility of significantly enhancing the Lee Teng program by incorporating a "hands-on" component to the students' summer research experience, making it a truly unique internship with the potential to attract future accelerator scientists at the critical undergraduate stage.

In addition to these laboratory-driven programs, Fermilab has also welcomed students from University groups carrying out research at Fermilab funded by grants to universities. Examples includes the development of photo-emission laser with University of Rochester, research on flat-beam generation with University of Chicago, or the investigation of beam-driven plasma wakefield acceleration with University of California, Los Angeles.

ASTA will be an excellent platform that could support and enhance the aforementioned education programs and support extramural research engaging students from Universities. Given the breath of the scientific program described in this proposal, we expect ASTA to provide support for topics of research in Advanced Accelerator R&D, laser science, beam diagnostics and synchronization, accelerator-based light source, superconducting RF system, and Beam Physics. Finally, some of the proposed topics could also foster research beyond Accelerator Science such as the utilization of single-cycle Terahertz pulses for solid-state Physics research or the use of X-rays for developing phase-contrast imaging (with possible spin-off to medical imaging).

ASTA will also offer a unique opportunities for the US Particle Accelerator School [8] sessions and classes. A number of hand-on, practical training laboratory sessions can use the ASTA accelerators – the photoinjector, the SRF cryomodule and the IOTA ring. Among them, those which were very popular in the past US PAS sessions:
  a. Modern RF systems class
  b. Beam Instrumentation Lab
  c. Fundamentals of Accelerator Instrumentation
  d. Beam Measurements and Diagnostics in Linacs and Rings
  e. Beam Dynamics Experiments at the IOTA ring
  f. Beam Measurements, Manipulation and Instrumentation in SRF Linac

Proximity of the Illinois Accelerator Research Center (IARC) at Fermilab [9] will make these classes attractive for the industrialists and technologists. At the IARC, scientists and engineers from Fermilab, Argonne and Illinois universities will work side by side with industrial partners to research and develop breakthroughs in accelerator science and translate them into applications for the nation's health, wealth and security. Besides a dedicated state-of-the-art facility that will house offices, technical and educational space to study cutting-edge accelerator technologies, it is anticipated that IARC's private industry partners will have access to the nearby ASTA facility to perform high power beam tests and collaborate with ASTA users from the US universities in training a new generation of scientists, engineers and technical staff in accelerator technology. These partnerships will make critical contributions to the technological and economic health of Illinois and the nation in general.